\documentclass[twocolumn]{aastex62}
\usepackage{natbib}
\usepackage{hyperref}
\newcommand{\uat}[2]{\href{http://astrothesaurus.org/uat/#2}{#1 (#2)}}

\newcommand\longname{PSO~J308.0416$-$21.2339} 
\newcommand\shortname{PJ\ 308$-$21}
\newcommand\luv{$L_{2500\, \textup{\footnotesize \AA}}$}
\graphicspath{{./}{figures/}}
\usepackage{graphics,graphicx}

\received{2019 September 18}
\accepted{2019 November 6}
\published{2019 December 18}

\submitjournal{the Astrophysical Journal}
\shorttitle{X-rays in a $z\sim6.2$ Quasar/Galaxy Merger}
\shortauthors{Connor et al.}

\begin{document}

\title{X-ray Observations of a $z\sim6.2$ Quasar/Galaxy Merger}

\correspondingauthor{Thomas Connor}
\email{tconnor@carnegiescience.edu}

\author[0000-0002-7898-7664]{Thomas Connor}
\affiliation{The Observatories of the Carnegie Institution for Science, 813 Santa Barbara St., Pasadena, CA 91101, USA}
\affiliation{Jet Propulsion Laboratory, California Institute of Technology, 4800 Oak Grove Dr., Pasadena, CA 91109, USA}

\author[0000-0002-2931-7824]{Eduardo Ba\~nados}
\affiliation{Max Planck Institute for Astronomy, K\"onigstuhl 17, D-69117 Heidelberg, Germany}
\affiliation{The Observatories of the Carnegie Institution for Science, 813 Santa Barbara St., Pasadena, CA 91101, USA}

\author[0000-0003-2686-9241]{Daniel Stern}
\affiliation{Jet Propulsion Laboratory, California Institute of Technology, 4800 Oak Grove Dr., Pasadena, CA 91109, USA}

\author[0000-0002-2662-8803]{Roberto Decarli}
\affiliation{INAF -- Osservatorio di Astrofisica e Scienza dello Spazio di Bologna, via Gobetti 93/3, I-40129, Bologna, Italy}

\author[0000-0002-4544-8242]{Jan-Torge Schindler}
\affiliation{Max Planck Institute for Astronomy, K\"onigstuhl 17, D-69117 Heidelberg, Germany}

\author[0000-0003-3310-0131]{Xiaohui Fan}
\affiliation{Steward Observatory, University of Arizona, 933 N. Cherry Ave., Tucson, AZ 85721, USA}

\author[0000-0002-6822-2254]{Emanuele Paolo Farina}
\affiliation{Max Planck Institut f\"ur Astrophysik, Karl-Schwarzschild-Stra{\ss}e 1, D-85748, Garching bei M\"unchen, Germany}

\author[0000-0002-5941-5214]{Chiara Mazzucchelli}
\affiliation{European Southern Observatory, Alonso de Cordova 3107, Vitacura, Region Metropolitana, Chile}

\author[0000-0003-2083-5569]{John S. Mulchaey}
\affiliation{The Observatories of the Carnegie Institution for Science, 813 Santa Barbara St., Pasadena, CA 91101, USA}

\author[0000-0003-4793-7880]{Fabian Walter}
\affil{Max Planck Institute for Astronomy, K\"onigstuhl 17, D-69117 Heidelberg, Germany}
\affil{National Radio Astronomy Observatory, Pete V. Domenici Array Science Center, P.O. Box O, Socorro, NM 87801, USA}

\begin{abstract}

Quasars at early redshifts ($z > 6$) with companion galaxies offer unique insights into the growth and evolution of the first supermassive black holes. Here, we report on a 150 ks \textit{Chandra} observation of \longname, a $z=6.23$ quasar with a merging companion galaxy identified in [\ion{C}{2}] and rest-frame UV emission. With  $72.3^{+9.6}_{-8.6}$ net counts, we find that \longname\ is powerful ($L_X = 2.31^{+1.14}_{-0.76} \times 10^{45}\ \textrm{erg}\,\textrm{s}^{-1}\,\textrm{cm}^{-2}$ in rest-frame $2.0\textrm{--}10.0$ keV) yet soft (spectral power-law index $\Gamma=2.39^{+0.37}_{-0.36}$ and optical-to-X-ray slope $\alpha_{\rm OX} = -1.41 \pm 0.11$). In addition, we detect three hard-energy photons $2\farcs0$ to the west of the main quasar, cospatial with the brightest UV emission of the merging companion. As no soft-energy photons are detected in the same area, this is potentially indicative of a highly-obscured source. With conservative assumptions, and accounting for both background fluctuations and the extended wings of the quasar's emission, these photons only have a probability $P=0.021$ of happening by chance. If confirmed by deeper observations, this system is the first high redshift quasar and companion individually detected in X-rays and is likely a dual AGN.

\end{abstract}


\section{Introduction} \label{sect:intro}
In recent years, the number of known quasars seen in the first billion years of the universe ($z\gtrsim5.7$) has exploded \citep[e.g.,][]{2015ApJ...801L..11V,2016ApJS..227...11B,2019AJ....157..236Y}, allowing new insights \citep[e.g.,][]{2017ApJ...840...24E,2019ApJ...884L..19D} into the populations of the earliest supermassive black holes (SMBHs). However, explaining the formation and initial evolution of these objects remains challenging \citep{2019ConPh..60..111S}. To this end, X-ray observations of these quasars are critical, as they provide the best view of the inner regions of the Active Galactic Nucleus (AGN) powering the quasar emission \citep{2016AN....337..375F}.

\begin{figure*}
\begin{center}
\includegraphics{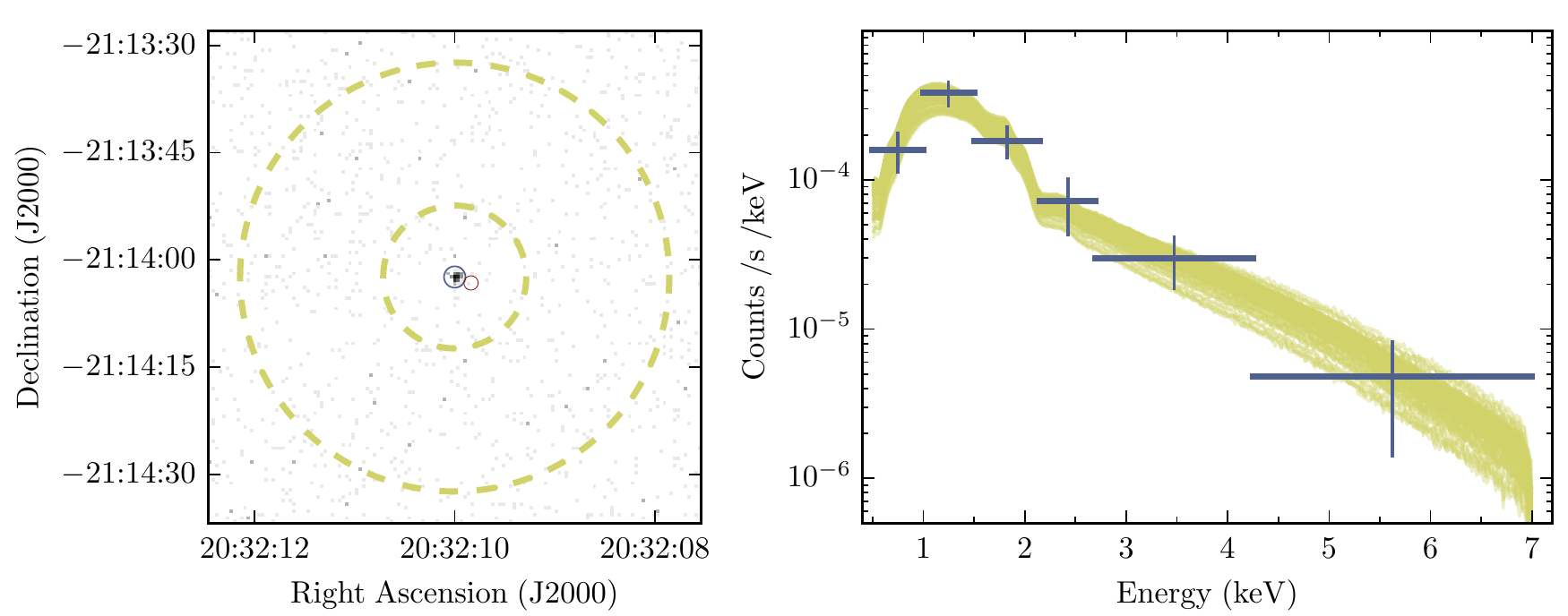}
\end{center}
\caption{{\bf Left}: \textit{Chandra} observation of \shortname, in the energy range $0.5-7.0$ keV. The image is binned to pixels of size $0.49\arcsec$, and the source and background regions are marked by a blue circle and yellow annulus, respectively. The quasar is strongly detected, and there are no other sources in the background region. We show in red the position of the candidate companion detection discussed in Section \ref{sect:Companion}. {\bf Right}: X-ray spectrum of \shortname\ (blue), with 100 well-fitting spectra underlaid (yellow). The spectrum is binned for ease of display, but fitting was performed on unbinned data. The 100 mock spectra are drawn from a Monte Carlo sampling of the fit, as described in the text.}
\label{fig:xray_image}
\end{figure*}

To date, most of the X-ray analyses of these quasars have focused on individual objects, and the current population of observed high-redshift quasars is both small and mostly only barely detected \citep{2017A&A...603A.128N,2019A&A...630A.118V}. Nevertheless, studies of individual quasars are still highly fruitful at characterizing mechanisms for early SMBH growth. For example, \citet{2018ApJ...856L..25B} showed that AGN are already X-ray luminous at $z=7.5$, \citet{2019A&A...628L...6V} identified a heavily obscured quasar candidate in a $z=6.5$ quasar/galaxy pair, and \citet{2018A&A...614A.121N} found potential evidence of AGN variability and of jets from a $z=6.3$ quasar. 

Of particular interest is \longname\ (hereafter \shortname), a quasar at $z=6.2341 \pm 0.0005$ discovered by \citet{2016ApJS..227...11B} and whose systemic redshift is accurately measured from the [\ion{C}{2}] emission of its host galaxy \citep{2018ApJ...854...97D}. With shallow ($\sim$10 minutes on-source) Atacama Large Millimeter Array (ALMA)  observations, \citet{2017Natur.545..457D} found that the quasar has a [\ion{C}{2}]-bright companion. Follow-up deeper and  higher-resolution observations with ALMA and the {\it Hubble Space Telescope} (\textit{HST}) by \citet{2019ApJ...880..157D} showed that the companion is visible on both sides of the quasar, spanning over $4\farcs0$ (20 kpc), and its kinematics can be explained by a toy model of a satellite galaxy being tidally stripped as it passes close to the quasar host galaxy. This makes \shortname\ one of the earliest galaxy mergers ever imaged \citep{2019ApJ...880..157D}.
\begin{figure*}
\begin{center}
\includegraphics{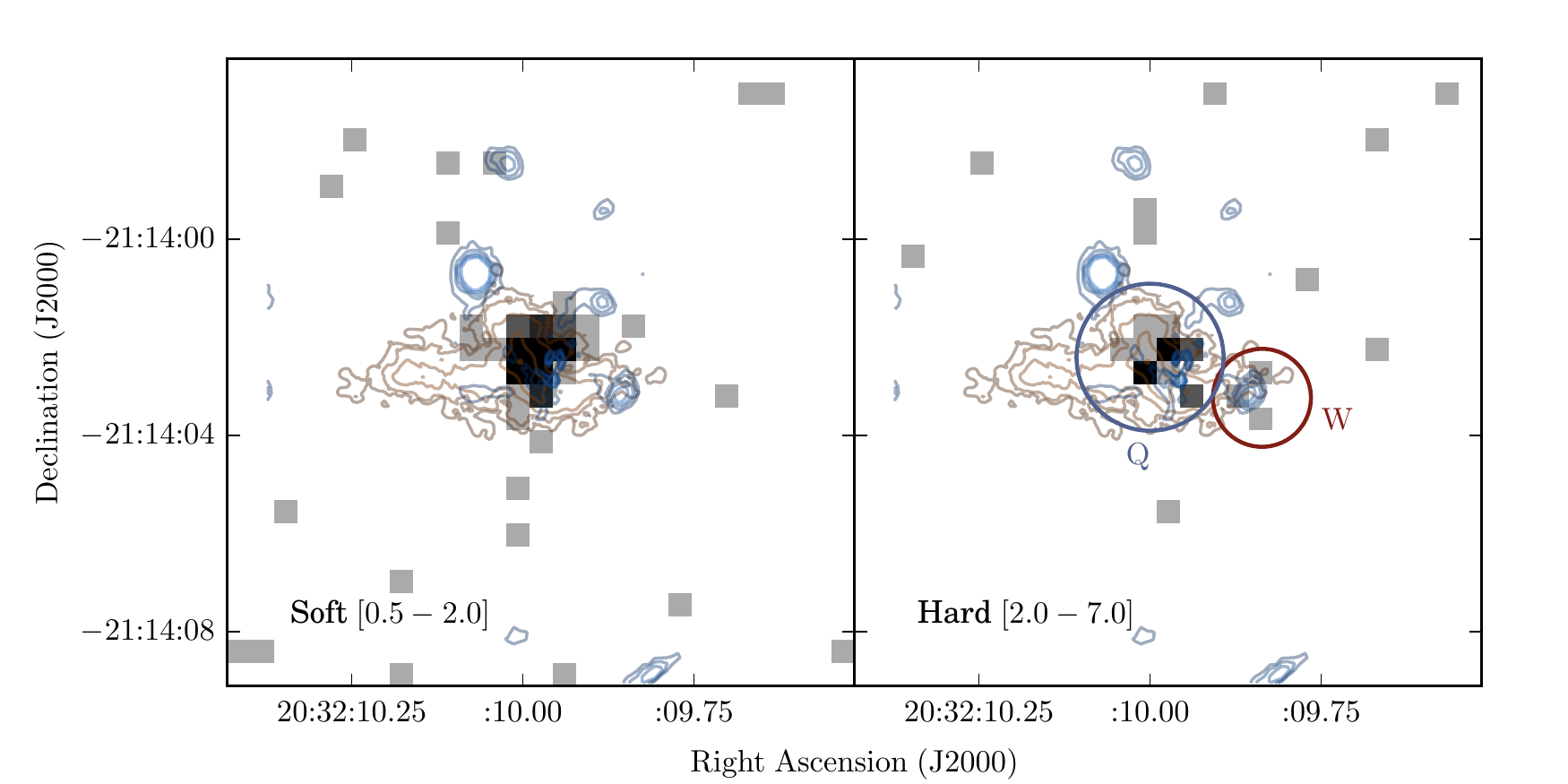}
\end{center}
\caption{Greyscale image of X-ray observations of \shortname, with ALMA [\ion{C}{2}] contours (yellow) and \textit{HST} rest-frame UV contours (blue, where the quasar has been removed through PSF subtraction) from \citet{2019ApJ...880..157D} overlaid. On the left and right, the quasar is shown in the soft (0.5--2.0 keV) and hard (2.0--7.0 keV) energy bands, respectively. A faint X-ray source to the west of the quasar is seen only in the hard band, coincident with the bright UV knot ``W'' reported by \citet{2019ApJ...880..157D}. \shortname\ (``Q'') and ``W'' are indicated by the blue and red circles, respectively. The bright object to the NE of the quasar in the \textit{HST} imaging is a foreground source.}\label{fig:sky_image}
\end{figure*}

In this paper, we describe X-ray observations of \shortname\ with {\it Chandra}, which are among the deepest X-ray observations of a $z>6$ quasar yet taken. One of the main objectives of our study was to see if the [\ion{C}{2}]-bright companion could host a faint AGN; given that the first SMBHs likely grew through galaxy mergers, the chances of finding lower-luminosity AGN are enhanced around the most distant quasars \citep{2016PASA...33...54R}. We describe our observations in Section \ref{sect:Obs} and report the X-ray properties of the optically selected quasar in Section \ref{sect:QSOprops}. In Section \ref{sect:Companion} we present a potential X-ray (heavily obscured) counterpart to the companion galaxy seen by ALMA and \textit{HST}. Finally, in Section \ref{sect:Disc} we discuss our results and implications.
We use a flat cosmology with $H_0 = 70\,\textrm{km\,s}^{-1}\,\textrm{Mpc}^{-1}$, $\Omega_M = 0.3$, and $\Omega_\Lambda = 0.7$. We assume a Galactic absorption column density toward \shortname\ of $N_{\rm H} = 4.02 \times 10^{20}\,\textrm{cm}^{-2}$ \citep{2005A&A...440..775K}. We adopt a quasar redshift of $z=6.234$; at this redshift, the scale is $5.59\,\textrm{kpc}\,\textrm{arcsec}^{-1}$. Errors are reported at the 1$\sigma$ (68\%) confidence level unless otherwise stated. Upper limits correspond to $3\sigma$ limits.

\section{Observations and Data Reduction}\label{sect:Obs}

We observed \shortname\ with the the Advanced CCD Imaging Spectrometer \citep[ACIS;][]{2003SPIE.4851...28G} on {\it Chandra} for a total of 150.92 ks as part of Sequence Number 703573. Observations were distributed across three visits: on 2018 August 27 (44.48 ks, Obs ID: 20470), 2018 August 29 (73.37 ks, Obs ID: 21725), and 2018 August 30 (33.07 ks, Obs ID: 21726). The detection image is shown in Figure \ref{fig:xray_image}. This is the second-deepest \textit{Chandra} observation of a $z\gtrsim6$ quasar \citep{2018A&A...614A.121N} and third-deepest observation of this high-redshift population with either \textit{Chandra} or \textit{XMM-Newton} (\citealt{2014MNRAS.440L..91P} and \citealt{2014A&A...563A..46M}; see \citealt{2019A&A...630A.118V} and \citealt{2019arXiv191004122P} for a full list).
Observations were conducted with the Very Faint telemetry format and the Timed Exposure mode, and {\it Chandra} was pointed so that \shortname\ fell on the ACIS-S3 chip. We analyzed these data using {\tt CIAO} version 4.11 \citep{2006SPIE.6270E..1VF} and CALDB version 4.8.2.  We used the ACIS standard filters for event grades (0, 2, 3, 4, and 6) and good time intervals. Observations were reduced with \texttt{chandra\_repro} with the parameter \texttt{check\_vf\_pha=yes} to reduce the quiescent background.

Observations were combined using the {\tt merge\_obs} routine to create images in the broad (0.5--7.0 keV) energy band for spatial analysis. The quasar is shown in the left panel of Figure \ref{fig:xray_image}. 
For spectral analysis, we used \texttt{specextract} to extract a source spectrum from $0.5-7.0$ keV ($3.6-50.6$ keV in the quasar rest-frame) within a circular region of radius $1\farcs5$ and a background from an annular region of inner radius $10\farcs0$ and width $30\farcs0$, both centered on \shortname. We detect $72.3^{+9.6}_{-8.6}$ background-subtracted net counts from the source in the 0.5-7.0 keV spectral range.
Spectral analysis was conducted using XSPEC v12.9.1p \citep{1996ASPC..101...17A} via the pyXspec utility. Due to the relatively low number of counts for Gaussian analysis (see, e.g., \citealt{2007A&A...468..501A} and \citealt{2009ApJ...693..822H} for further discussion), we did not bin our spectra, and used the modified C-statistic \citep{1979ApJ...228..939C, 1979ApJ...230..274W} to determine the best-fitting parameters. The quasar spectrum was modeled as a power law with Galactic dust absorption, using the xspec models \texttt{phabs}$\times$\texttt{powerlaw}, and with rest-frame dust absorption included, using the models \texttt{zphabs}$\times$\texttt{phabs}$\times$\texttt{powerlaw}. The binned spectrum is shown in Figure \ref{fig:xray_image}.

\section{Properties of \shortname}\label{sect:QSOprops}
We first quantified the X-ray hardness ratio\footnote{$\mathcal{HR} = (H-S)/(H+S)$, where $H$ and $S$ are the net counts in the hard (2.0--7.0 keV) and soft (0.5--2.0 keV) bands, respectively.} using the Bayesian techniques described by \cite{2006ApJ...652..610P}. We assumed uniform (Jeffreys) priors and integrated the posterior distribution with Gaussian quadrature. The hardness ratio of \shortname\ is $\mathcal{HR} = -0.48^{+0.11}_{-0.10}$ across the observed 0.5--2.0 keV and 2.0--7.0 keV bands. The quasar spectrum is fairly soft, in agreement with broader population trends reported by \citet{2017A&A...603A.128N}. Soft- and hard-band images of \shortname\ are shown in Figure \ref{fig:sky_image}.

\begin{figure*}
\begin{center}
\includegraphics{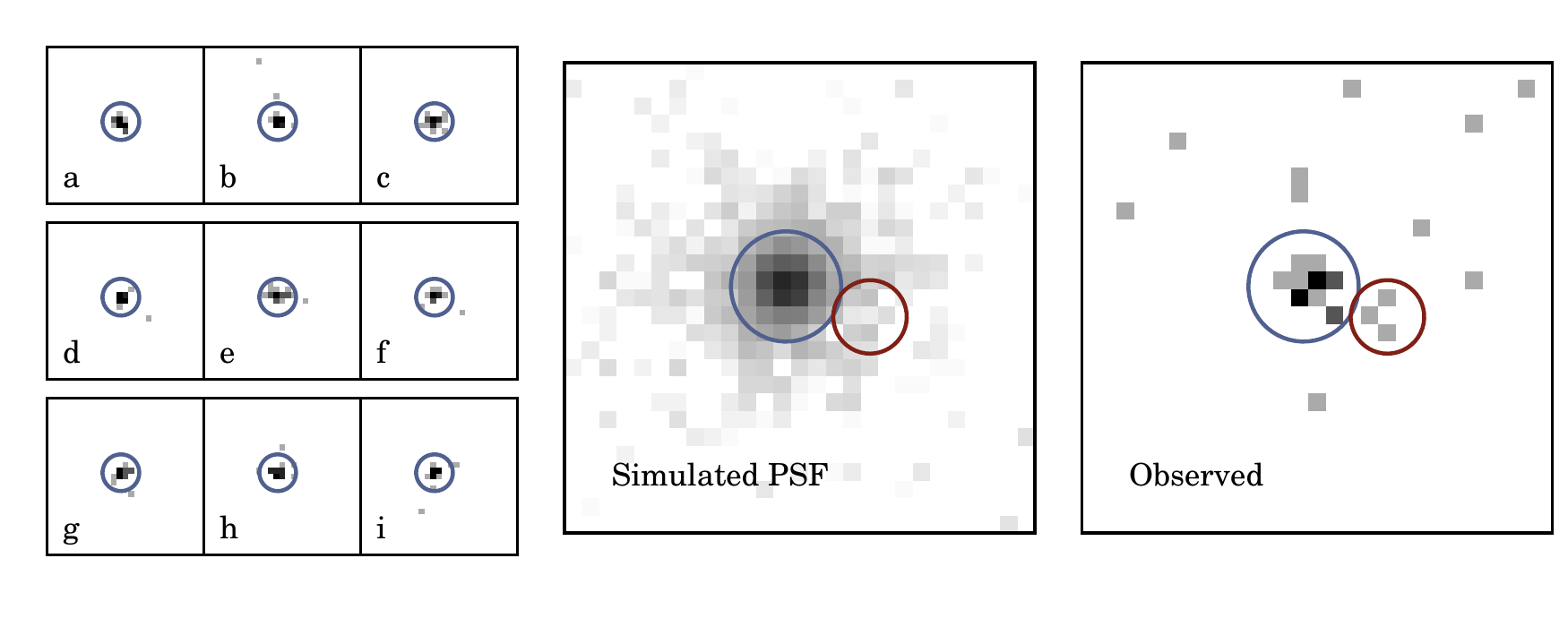}
\end{center}
\caption{Comparison between the expected PSF of a single source in hard energies to the observed structure. \textbf{Left:} nine simulated PSFs, as described in the text, showing the same number of counts as are seen for the quasar and companion. \textbf{Center:} the summed PSF from 1000 simulations. \textbf{Right:} the observed hard-energy X-ray flux around \shortname. All images cover the same region shown in Figure \ref{fig:sky_image}, and the blue and red circles show the same regions shown in that figure. The excess structure observed to the west is not consistent with being produced by the PSF alone. Note that in the central panel flux is scaled logarithmically over a larger range than in the right panel to show the extended wings of the PSF.}\label{fig:PSF_sims}
\end{figure*}

From the best-fit of the quasar spectrum, we find that the power-law index $\Gamma=2.39^{+0.37}_{-0.36}$ and the luminosity from 2.0 to 10.0 keV (rest-frame) is $L_X = 2.31^{+1.14}_{-0.76} \times 10^{45}\ \textrm{erg}\,\textrm{s}^{-1}\,\textrm{cm}^{-2}$. Errors were computed by evaluating the Cash statistic, $C$, across the distribution of model parameters for \texttt{powerlaw}; uncertainties on $\Gamma$ include all values where offsets from the best-fitting Cash statistic $\Delta C \leq 2.30$ \citep[two free parameters, e.g.,][]{1976ApJ...208..177L}. For $L_X$, model luminosities were computed for every set of parameters, and the reported uncertainties include all models with $\Delta C \leq 2.30$. As a cross-check, we also produced 100 fake spectra using the XSPEC \texttt{fakeit} command drawn from a Monte Carlo sampling of the spectral fit; these are shown in Figure \ref{fig:xray_image}.

We also consider the case where there is absorption at the redshift of the quasar. To do this, we modeled the emission with the xspec models \texttt{phabs}$\times$\texttt{zphabs}$\times$\texttt{powerlaw}. Here, \texttt{phabs} is left to the Galactic value, but \texttt{zphabs} is set to redshift $z=6.234$ with absorbing column density $N_{H,z}$ allowed to vary. With three free parameters, the errors include all values with $\Delta C\leq3.53$. For this model, we find $\Gamma=2.8^{+1.1}_{-0.7}$, $L_X = 4.9^{+25.5}_{-3.3} \times 10^{45}\ \textrm{erg}\,\textrm{s}^{-1}\,\textrm{cm}^{-2}$ and $N_{H,z} = 2.4^{+5.5}_{-2.4}\times\,10^{24}\,\textrm{cm}^{-2}$. Unsurprisingly, $N_{H,z}$ is not well constrained, as in the high rest-frame energies we probe, variations in the spectral shape are minimal for column densities of $N_{H,z} < 10^{24}\,\textrm{cm}^{-2}$. As including the extra term provides no meaningful constraints, and since the presence of broad lines in the optical spectrum implies a column density of $N_{H,z}\lesssim\,10^{22}\,\textrm{cm}^{-2}$ -- which our spectrum is not sensitive to -- we  hereafter only consider the results when \texttt{zphabs} is not included.

Finally, we consider the X-ray-to-optical power-law slope, $\alpha_{\rm OX},$ defined as  
\begin{equation}
\alpha_{\rm OX}= 0.3838 \times \log( {L}_{2\,\mathrm{keV}} / {L}_{2500\,\textup{\footnotesize \AA}}),
\end{equation}
where $L_{2\,\mathrm{keV}}$ and \luv\ are the monochromatic luminosities at rest-frame 2 keV and 2500\AA, respectively. We compute \luv\ from the previously reported value of $m_{1450} = 20.46$ \citep{2016ApJS..227...11B}; to make this conversion, we assume that flux in this spectral region scales as $f_{\nu} \propto \nu^{\alpha_\nu}$ and adopt $\alpha_\nu=-0.3$, the value used to compute $m_{1450}$ by \citet{2016ApJS..227...11B}. We find $\alpha_{\rm OX} = -1.41 \pm 0.11$.

To provide further context for the properties of \shortname, we also report here measurements of the quasar's black hole mass and Eddington ratio, taken from an upcoming analysis of high-redshift quasars (Farina et al.\ and Schindler et al.\ in preparation); these results come from single-epoch virial estimators derived using the \ion{Mg}{2} line detected in the near-infrared spectrum. They find a black hole mass of $1\times 10^9\,M_\odot$ and an Eddington ratio of $1.5$, where the statistical uncertainties are dwarfed by the roughly $0.55\,\rm{dex}$ systematic uncertainties on both quantities inherent in this technique \citep{2004ApJ...615..645O,2006ApJ...641..689V}.

\section{Potential Companions}\label{sect:Companion}

One reason \shortname\ was observed was to look for X-ray counterparts to the companion first observed by \citet{2017Natur.545..457D} and later by \citet{2019ApJ...880..157D}. In Figure \ref{fig:sky_image} we show the area around \shortname\ in the soft- and hard-energy bands, with the [\ion{C}{2}] emission contours and \textit{HST} rest-frame UV contours from \citet{2019ApJ...880..157D} overplotted. While there does not appear to be any structure to the east of the quasar, three closely spaced hard X-ray photons were detected just to the west of \shortname, coincident with a bright knot of stellar light observed by \textit{HST}. These photons align with the outer extent of the [\ion{C}{2}] emission and have no corresponding soft-band photons, highly suggestive of a heavily obscured quasar. Below, we discuss these results. 

First, however, we quantify the astrometric alignment between these {\it Chandra} observations and the previous {\it HST} and ALMA observations. Using \texttt{WAVDETECT} \citep{2002ApJS..138..185F}, we identify the positions of sources in our broadband image. Comparing these positions with the Guide Star Catalog v2.3 \citep{2008AJ....136..735L} using \texttt{wcs\_match}, the average residual is $0\farcs5$, which is the size of ACIS pixels. Likewise, we compare the measured centroid of \shortname\ to the ALMA coordinates reported by \citet{2017Natur.545..457D}. Again, we find an offset of $\lesssim0\farcs5$. As can be seen in Figure \ref{fig:sky_image}, even if our astrometry is off by $0\farcs5$, a shift of this level will not produce a significant effect on the results discussed below.

We therefore begin our analysis with the three photons to the west. To estimate the probability that these photons arise solely from the background, we use binomial statistics to calculate probabilities, as described by \citet{2007ApJ...657.1026W} and \citet{2014ApJ...785...17L}. We counted three photons within a standard $1\farcs0$ aperture in the traditional ACIS hard-energy range \citep[$2.0-7.0$\ keV, e.g.,][]{2010ApJS..189...37E}; we also extracted a background centered on our target in the same annulus used for spectroscopic analysis, finding 380 counts in the $10\arcsec - 30\arcsec$ aperture. We find the probability of these sources arising from the background alone to be $P = 0.013$. Similarly, we randomly placed 10,000 $1\farcs0$ radius apertures in the vicinity of \shortname, excluding the $5\farcs0$ closest to the quasar, and counted the number of apertures that have at least three hard X-ray counts; from this, we estimate the probability of three counts in this small of an aperture to be $P = 0.013$.
\begin{figure*}
\begin{center}
\includegraphics{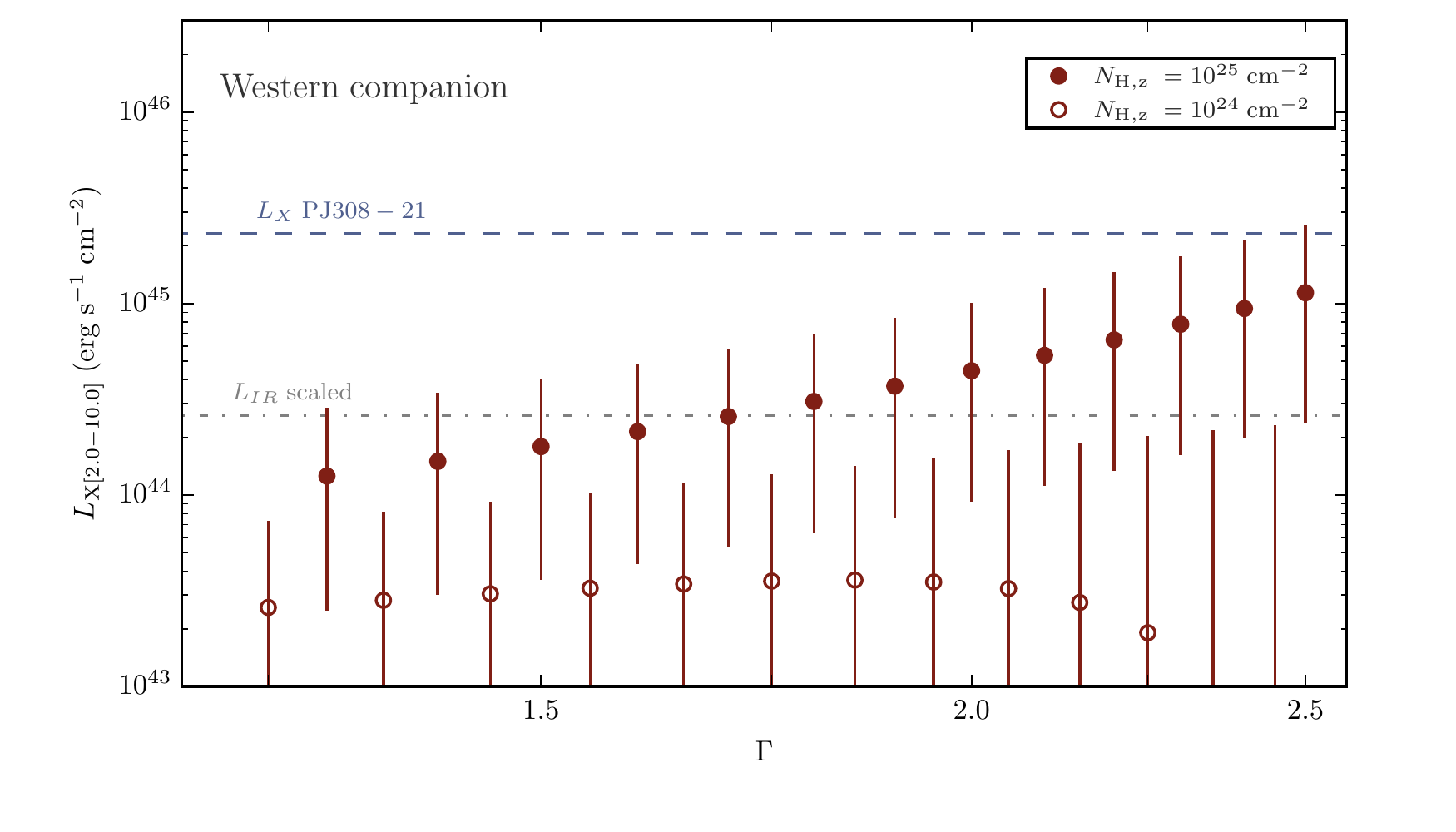}
\end{center}
\caption{Best-fit luminosities with $1\sigma$ uncertainties on the unabsorbed rest-frame luminosity of the candidate obscured detection to the west of \shortname\ for two assumed values of rest-frame column density.  The X-ray luminosity of \shortname\ is indicated by the dashed line. We also scale that value by the IR luminosities reported by \citet{2019ApJ...880..157D} for the companion to provide a rough idea of its expected X-ray luminosities; this scaled value is marked by the dotted-dashed line. If the western region of the companion is host to a quasar, that quasar must be some combination of X-ray-faint, heavily dust-obscured, and/or have a steep power-law index.}\label{fig:companion_limits}
\end{figure*}

Of course, the emission can also be influenced by the presence of the nearby X-ray bright source, \shortname. In all three observations, \shortname\ was observed almost on-axis, such that the expected point-spread function (PSF) should be small. To  better understand the expected behavior of the {\it Chandra} PSF, we simulated 1000 observations of a monochromatic 4 keV point source located at the position of \shortname\ using the \textit{Chandra} Ray Tracer (ChaRT). These simulations provide tens of thousands of simulated photons to quantify the odds of a given photon appearing in the western region. The ChaRT output was in turn processed by marx v 5.4.0 \citep{2012SPIE.8443E..1AD} to generate synthetic PSF images. Nine of these images, which have been capped to only show the same number of counts as seen for \shortname\ and the candidate western companion, are shown in Figure \ref{fig:PSF_sims}, as is the full combination of all 1000 simulations. As shown in that figure, the three counts do not normally arise from the PSF; the expectation is for the PSF to only contribute 0.108 counts to the site of the potential companion, based on the flux in the inner $1\farcs5$. Combining the effects of the PSF and the background, the probability of these three counts appearing by chance is still only $P= 0.021$.

It should be noted that we have adopted several conservative restrictions, and that by relaxing these the statistical significance could be increased -- notably a more restricted energy range. Additionally, a fourth hard-energy photon in the companion aperture was excised through the choice to set \texttt{check\_vf\_pha=yes} in the original reprocessing. Use of this mode is cautioned, as it may reject real events near bright sources\footnote{See \url{http://cxc.harvard.edu/ciao/why/aciscleanvf.html\#real\_events}}. However, even with these conservative choices, these photons can only be explained by a random fluctuation with probability $ P=0.021$, and this detection is coincident not only with the blue-shifted [\ion{C}{2}] emission reported by \citet{2019ApJ...880..157D} but also with the brightest knot of starlight identified in that work in \textit{HST} imaging. 

To place limits on the potential companion, we fit the observed $0.5 - 7.0$ keV spectrum with \texttt{XSPEC}, using a \texttt{phabs}$\times$\texttt{zphabs}$\times$\texttt{powerlaw} model. We fixed the redshift and Galactic $N_H$, and evaluated the model for a series of fixed values of $\Gamma$ and redshifted column density, $N_{H,z}$. The corresponding unobscured rest-frame $2.0-10.0$ keV luminosities that best fit the potential companion are shown in Figure \ref{fig:companion_limits}. In addition, we also mark the observed luminosity for \shortname\ and a predicted luminosity for the companion in the simplistic case that X-ray and IR luminosities are proportional. The fits show that if these photons are produced by a quasar, it must be some combination of highly shrouded, low-luminosity, or soft. We also note that, as we did not detect any photons from 0.5 to 7.0 keV for the eastern companion, any potential X-ray source associated with that region must be even fainter.

\section{Discussion}\label{sect:Disc}

While the average value of the X-ray spectral power-law index, $\Gamma$, for $z > 5.7$ quasars is $\Gamma\sim1.9$ \citep{2017A&A...603A.128N}, we found the spectrum of \shortname\ was best fit with a softer value, $\Gamma=2.39^{+0.37}_{-0.36}$, which is one of the softest values for $z>6$ quasars yet found \citep[only considering those with reasonable constraints;][]{2019A&A...630A.118V}. A number of recent works have found a correlation between $\Gamma$ and the Eddington ratio (e.g., \citealt{2008ApJ...682...81S}, \citealt{2013MNRAS.433.2485B}, \citealt{2016ApJ...826...93B}; this is not without some controversy though, e.g., \citealt{2017MNRAS.470..800T}, \citealt{2019MNRAS.487.2463W}). Similarly, simulations \citep[e.g.,][]{2006ApJS..163....1H} predict that major mergers drive high accretion rates in AGN. Therefore, since this system is actively merging \citep{2019ApJ...880..157D} and accreting above the Eddington limit, the high value of $\Gamma$ might be expected.

We also report on a potential detection of the gas-rich companion as a heavily obscured X-ray source. While only three counts are observed in hard X-rays, they are coincident with the extended [\ion{C}{2}] emission and the brightest rest-frame UV knot reported by \citet{2019ApJ...880..157D}. Recently, \citet{2019A&A...623A.172C} showed that the host galaxies of $z>2.5$ AGN can provide significant levels of obscuration; from the toy model for this system presented by \citet{2019ApJ...880..157D}, we expect to be looking through the edge-on leading edge of the companion galaxy. Adopting the simple model that the entire gas mass reported by \citet{2019ApJ...880..157D} for the western companion is composed of hydrogen, and adopting the \citet{2019A&A...623A.172C} toy model that all the gas is in a uniform-density sphere of half-mass radius $r_H= 1.0\ \textrm{kpc}$, then the total column density to a source at the center of this sphere is $N_{H,z} \approx 4 \times 10^{23}\ \textrm{cm}^{-2}$. We note both that the gas mass was measured over a larger radius than 1 kpc and that it is clearly non-uniform in density, so that this column density is at best an order-of-magnitude estimate, or even an upper limit. Nevertheless, the observed gas mass could, depending on its structure and orientation, potentially contribute to the column density needed to to explain the lack of observed soft X-ray photons.

It is also worth considering \shortname\ in the context of PSOJ167$-$13 ($z=6.5$). Both quasars have [\ion{C}{2}] bright companions \citep{2017Natur.545..457D, 2017ApJ...850..108W} reminiscent of ongoing mergers, and these are the only systems at redshifts $z>6$ with companions detected in [\ion{C}{2}] and in rest-frame UV \citep{2019ApJ...880..157D, 2019ApJ...881..163M,2019ApJ...882...10N}. While here we report that \shortname\ is X-ray luminous and its companion may host a heavily obscured X-ray source, in an analysis of PSOJ167$-$13 ($z=6.5$) \citet{2019A&A...628L...6V} reported a detection of only one source, which also appears to be heavily obscured. That these two quasars beyond $z>6$ have both rest-frame UV-bright companions and heavily obscured quasar candidates may be consistent with the prediction of \citet{2019A&A...623A.172C}, that the host's interstellar medium is capable of producing significant column densities of $N_{H}$ around high-redshift quasars.

The results presented here required 150 ks of observations with \textit{Chandra}; despite this, the potentially detected companion can only be ruled out as a fluctuation to probability $P=0.021$. This work, as well as that of \citet{2019A&A...628L...6V}, \citet{2018ApJ...856L..25B}, and \citet{2018A&A...614A.121N}, pushes the current generation of X-ray telescopes to their limits. Future missions with high resolution and/or improved collecting areas, such as \textit{Lynx} \citep{2019JATIS...5b1001G}, \textit{AXIS} \citep{2019arXiv190304083M}, and {\it Athena} \citep{2013arXiv1306.2307N}, will be necessary to advance our understanding of the earliest quasars.
\acknowledgments

{\small We thank the anonymous referee for their productive and helpful comments. T.C. was supported by STScI/NASA award HST-GO-15198. The work of T.C. and D.S. was carried out at the Jet Propulsion Laboratory, California Institute of Technology, under a contract with NASA. The scientific results reported in this article are based on observations made by the \textit{Chandra X-ray Observatory}. This research has made use of software provided by the \textit{Chandra} X-ray Center (CXC) in the application package CIAO.
This work is based on observations made with the NASA/ESA \textit{Hubble Space Telescope}, obtained from the Data Archive at the Space Telescope Science Institute, which is operated by the Association of Universities for Research in Astronomy, Inc., under NASA contract NAS 5-26555. 
These observations are associated with program 14876. Support for this work was provided by NASA through grant number 10747 from the Space Telescope Science Institute, which is operated by AURA, Inc., under NASA contract NAS 5-26555. This work is based on observations collected at the European Organisation for Astronomical Research in the Southern Hemisphere under ESO programme 098.B-0537(A).}

\facility{CXO, HST, ALMA, VLT:Kueyen}
\software{CIAO \citep{2006SPIE.6270E..1VF},
          MARX \citep{2012SPIE.8443E..1AD},
          XSPEC \citep{1996ASPC..101...17A}}

\bibliography{bibliography}

\end{document}